\documentclass[twocolumn,prd,aps,floats,showpacs]{revtex4}
\def\DESepsf(#1 width #2){\epsfxsize=#2 \epsfbox{#1}}
\usepackage{graphicx}
\usepackage{color}
\usepackage{amsmath}
\def\bmatrix{\left[\begin{array}}
\def\ematrix{\end{array}\right]}

\begin{document}

%

\let\a=\alpha      \let\b=\beta       \let\c=\chi        \let\d=\delta
\let\e=\varepsilon \let\f=\varphi     \let\g=\gamma      \let\h=\eta
\let\k=\kappa      \let\l=\lambda     \let\m=\mu
\let\o=\omega      \let\r=\varrho     \let\s=\sigma
\let\t=\tau        \let\th=\vartheta  \let\y=\upsilon    \let\x=\xi
\let\z=\zeta       \let\io=\iota      \let\vp=\varpi     \let\ro=\rho
\let\ph=\phi       \let\ep=\epsilon   \let\te=\theta
\let\n=\nu
\let\D=\Delta   \let\F=\Phi    \let\G=\Gamma  \let\L=\Lambda
\let\O=\Omega   \let\P=\Pi     \let\Ps=\Psi   \let\Si=\Sigma
\let\Th=\Theta  \let\X=\Xi     \let\Y=\Upsilon

%

%

\def\cA{{\cal A}}                \def\cB{{\cal B}}
\def\cC{{\cal C}}                \def\cD{{\cal D}}
\def\cE{{\cal E}}                \def\cF{{\cal F}}
\def\cG{{\cal G}}                \def\cH{{\cal H}}
\def\cI{{\cal I}}                \def\cJ{{\cal J}}
\def\cK{{\cal K}}                \def\cL{{\cal L}}
\def\cM{{\cal M}}                \def\cN{{\cal N}}
\def\cO{{\cal O}}                \def\cP{{\cal P}}
\def\cQ{{\cal Q}}                \def\cR{{\cal R}}
\def\cS{{\cal S}}                \def\cT{{\cal T}}
\def\cU{{\cal U}}                \def\cV{{\cal V}}
\def\cW{{\cal W}}                \def\cX{{\cal X}}
\def\cY{{\cal Y}}                \def\cZ{{\cal Z}}
%

\newcommand{\Ns}{N\hspace{-4.7mm}\not\hspace{2.7mm}}
\newcommand{\qs}{q\hspace{-3.7mm}\not\hspace{3.4mm}}
\newcommand{\ps}{p\hspace{-3.3mm}\not\hspace{1.2mm}}
\newcommand{\ks}{k\hspace{-3.3mm}\not\hspace{1.2mm}}
\newcommand{\des}{\partial\hspace{-4.mm}\not\hspace{2.5mm}}
\newcommand{\desco}{D\hspace{-4mm}\not\hspace{2mm}}
\renewcommand{\figurename}{Fig.}


%
\title{\boldmath
 CP Violation 
 in Fourth Generation Quark Decays}
\vfill
\author{Abdesslam Arhrib$^{1,2}$ and Wei-Shu Hou$^{1,3}$ \\
$^{1}$National Center for Theoretical Sciences,
National Taiwan University, Taipei, Taiwan 10617\\
$^{2}$Facult\'e des Sciences et Techniques, B.P 416 Tangier, Morocco\\
$^{3}$Department of Physics, National Taiwan University, Taipei,
Taiwan 10617 }
%

\date{\today}
%
%
%
\begin{abstract}
We show that, if a fourth generation is discovered at the Tevatron
or LHC, one could study $CP$ violation in $b' \to s$ decays.
Asymmetries could reach 30\% for $b'\to sZ$ for $m_{b'} \lesssim
350$ GeV, while it could be greater than 50\% for $b'\to s\gamma$
and extend to higher $m_{b'}$. Branching ratios are
$10^{-3}$--$10^{-5}$, and CPV measurement requires tagging. Once
measured, however, the CPV phase can be extracted with little
theoretical uncertainty.
\end{abstract}
\pacs{
 11.30.Er, 
 12.15.Ff, 
 12.15.Lk, 
 12.15.Mm  
 }
%
\maketitle

\pagestyle{plain}


%
Measurements of the phase angle $\sin2\beta/\phi_1 \equiv
\sin2\Phi_{B_d}$ in $B_d \to J/\psi K^0$ and other decays are in
good agreement~\cite{PDG} with the Kobayashi--Maskawa (KM)
model~\cite{KM}. This Standard Model (SM) with 3 generations
predicts $\sin2\Phi_{B_s}^{\rm SM} \simeq -0.04$~\cite{PDG} for
time-dependent CP violation (TCPV) in $B_s \to J/\psi \phi$ mode,
which is beyond the sensitivities at the Tevatron, and accessible
only by the LHCb experiment. Any indication for a finite value at
the Tevatron implies physics beyond SM (BSM).

Interestingly, both the CDF and D$\emptyset$ experiments
reported~\cite{CDFsin2betas} recently a large and negative value
for $\sin2\Phi_{B_s}$.
Though not yet significant, the central value echoes the
predictions~\cite{HNS} based on a fourth generation explanation of
the direct CPV (DCPV) difference, $\Delta A_{K\pi} \equiv
{A}_{B^+\to K^+\pi^0} - {A}_{B^0\to K^+\pi^-} \sim
+15\%$~\cite{PDG,BelleNature}, observed by the B factories.
By correlations of the $b\to s$ $Z$--penguin and $b\bar s
\leftrightarrow s\bar b$ box diagrams, a sizable 4th geneation
contribution to $\Delta A_{K\pi}$ would imply prominent CPV in
$B_s \to J/\psi \phi$. With $B_s$ mixing observed by CDF in 2006,
a stronger prediction was made. New results on $\sin2\Phi_{B_s}$
($\equiv -\sin2\beta_{s}$ of CDF) are eagerly awaited. To up the
ante, a 4th generation could enhance the invariant CPV measure of
Jarlskog~\cite{Jarlskog} by a factor
of $10^{15}$~\cite{Hou08}, and perhaps 
could satisfy the CPV part of the Sakharov
conditions~\cite{Sakharov} for baryogenesis in the early Universe.

The 4th generation is troubled by the electroweak precision test
(EWPT) $S$ parameter~\cite{Nnu}.
However, 
this severe constraint~\cite{PDG} is softened when one allows some
$t'$--$b'$ mass splitting that contributes to $T$
parameter~\cite{KPST07,4State}. With the LHC, we finally  have a
machine that can discover or rule out the 4th generation once and
for all by direct search~\cite{AH06}. There is in fact renewed
interest at the Tevatron. CDF has recently searched for $t'\to qW$
(no $b$-tagging) using 2.8 fb$^{-1}$, and for same sign
dileptons~\cite{CMSEXO} in $b'\bar b' \to t\bar
tW^-W^+$~\cite{AH06} based on 2.7 fb$^{-1}$ data, setting the 95\%
C.L. limits of $m_{t'} > 311$ GeV~\cite{tpCDF08} and $m_{b'}
> 325$ GeV~\cite{bpCDF09}, respectively. Note that these limits
assume predominance of the search mode. The $t'$ and especially
the $b'$ decays could be richer.

\begin{figure}[b!]
\vskip-6.2cm \hskip1.5cm
\includegraphics[height=13cm]{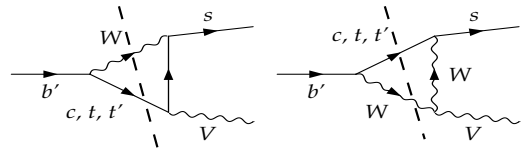}
 \vskip-5.2cm
\caption{
 $b'\to s$ loop transition, where the cut is for $c$ and
 $t$. 
 }
\label{fig:bptos}
\end{figure}

Motivated by the BSM source of CPV, and with heightened direct
search activity, we ask: What can the direct discovery of the $b'$
and $t'$ quarks do for the quest of CPV?
We find the best case to be flavor changing neutral current (FCNC)
$b'\to s$ decays, with $b'$ around and above the $tW$ threshold
(see Fig.~\ref{fig:bptos}).

The study of CPV in $b'\to s$ transitions complements the
traditional agenda of BSM CPV search in the flavor sector, such as
$B_s \to J/\psi \phi$, $K_L\to \pi^0\bar\nu\nu$, or $D^0$ mixing.
It opens up the new chapter of very heavy flavor CPV studies.
Interestingly, though of DCPV type
(there is no longer the exquisite TCPV mechanism for $b'$), the
$CP$ conserving phases in $b'\to s$ transitions are calculable.
Once measured, the CPV phase can in principle be extracted with
little theoretical uncertainty.


CPV requires two interfering amplitudes ${\cal M} = {\cal M}_1 +
{\cal M}_2$,
%
%
and the $CP$ asymmetry is
\begin{eqnarray}
A_{\rm{CP}}
 &=& \frac{2|{\cal M}_1||{\cal M}_2| \sin\delta \sin\phi}
          {|{\cal M}_1|^2 + |{\cal M}_2|^2
         + 2|{\cal M}_1||{\cal M}_2| \cos\delta \cos\phi},\;\
 \label{eq:cpa}
\end{eqnarray}
which vanishes unless both the weak {\it and} ``strong" phase
differences $\phi \equiv \phi_2-\phi_1$ and $\delta \equiv
\delta_2-\delta_1$ are nonzero.
%
%
In TCPV studies, $\delta = \Delta m_B \Delta t$ is measured,
allowing extraction of CPV phase $\phi$. For very heavy quarks, we
no longer expect meson formation because of rapid quark decay.
What is left is DCPV. Fortunately, unlike $B$ meson decays,
the absorptive amplitudes of Fig.~\ref{fig:bptos} are calculable,
and QCD corrections perturbative.

The $u$ quark effect in Fig.~\ref{fig:bptos} is suppressed by a
tiny $V_{ub}^*V_{ub'}$. Alternatively, $q = c,$ $u$ are
effectively massless, and provide a GIM subtraction to the $t$ and
$t'$ amplitudes via $V_{qs}^*V_{qb'} + V_{ts}^*V_{tb'} +
V_{t's}^*V_{t'b'} = 0$, giving
\begin{eqnarray}
{\cal M}_{b'\to s}
  =  V_{ts}^*V_{tb'}\,\Delta(t,0)
  +  V_{t's}^*V_{t'b'}\,\Delta(t',0),
 \label{eq:Mbptos}
\end{eqnarray}
with $\Delta(a,b) = f(m_a) - f(m_b)$, and $f$ the loop function.

Let us consider the $CP$ conserving phase difference $\delta$
between $\Delta(t,0)$ and $\Delta(t',0)$.
%
Having both $t'$, $b'$, and even the top rather heavy, enriches
the strong phase difference $\sin\delta$ in Eq.~(\ref{eq:Mbptos}).
For $b'$ below $tW$ (hence $t'W$) threshold, both the $t$ and $t'$
effects are dispersive,
the strong phase difference between GIM-subtracted $t$ and $t'$
contributions are subdued. But as the $tW$ threshold (illustrated
by the cut in Fig.~\ref{fig:bptos}) is approached, the dispersive
$t$ amplitude gets affected, and $\sin\delta$ starts to grow.
Above $tW$ threshold, the behavior of $\sin\delta$ depends on
$m_{t'}$, e.g. whether it is correlated with $m_{b'}$ due to
electroweak constraints. It also depends on the process.

\begin{figure}[t!]
\begin{tabular}{rr}
\hspace{-.3cm} \resizebox{44mm}{!}{\includegraphics{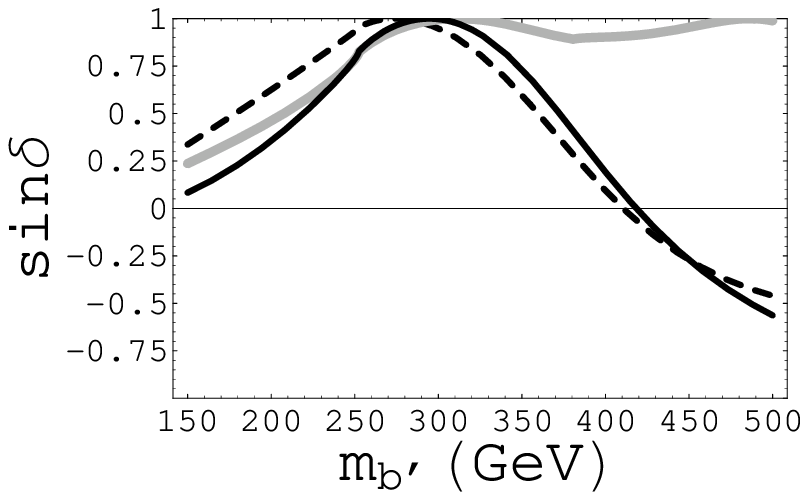}} &
\hspace{-.55cm} \resizebox{44mm}{!}{\includegraphics{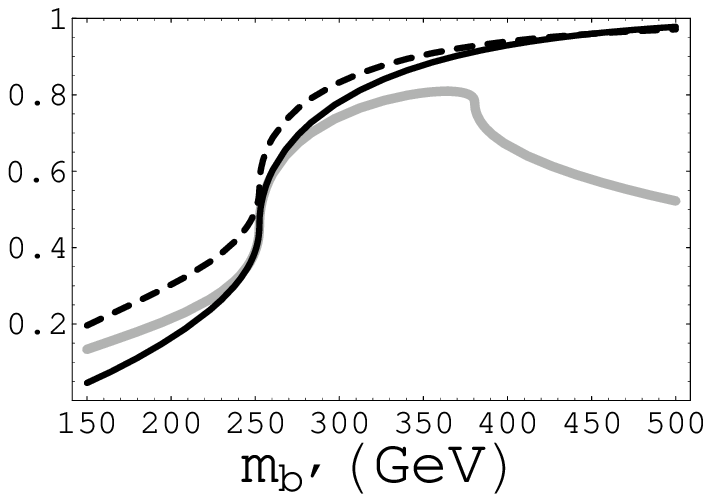}}
\end{tabular}
\caption{$CP$ conserving phase difference $\sin\delta$ between
$\Delta(t,0)$ and $\Delta(t',0)$ of Eq.~(\ref{eq:Mbptos}) as a
function of $m_{b'}$, for $b'\to sZ_L$ (left), where $Z_L$ is the
longitudinal $Z$, and $b' \to s\gamma$ (right). Solid line is for
$m_{t'}=m_{b'}+50$ GeV, and grey solid (dashed) line is for
$m_{t'}=300$ (500) GeV.} \label{fig:sindelta}
\end{figure}
We plot $\sin\delta$ vs. $m_b'$ in Fig.~\ref{fig:sindelta} for
three $t'$ mass scenarios, where 2(a) is for longitudinal $Z$
emission, and 2(b) for $b' \to s\gamma$ (the transverse $Z$
contribution to $b'\to sZ$ is similar). For the solid line of
$m_{t'}=m_{b'}+50$ GeV that satisfies the EWPT $S$--$T$
constraints, $\sin\delta$ rises to 1 for $b'\to sZ_L$ as one
crosses the $tW$ threshold, and stays close to 1 until it starts
to drop for $m_{b'} \gtrsim 310$ GeV, changing sign for $m_{b'}
\gtrsim 420$ GeV. A similar effect is seen for fixed $m_{t'} =$
500 GeV (dashed line),
where $\sin\delta$ is larger than the previous case for $b'$ below
$tW$ threshold. Thus, the flipping of sign of $\sin\delta$ is due
to not crossing the $t'W$ threshold. The behavior for $b'\to
s\gamma$ is different because it is a conserved current.

To illustrate $t'W$ threshold crossing, the grey solid line in
Fig.~\ref{fig:sindelta} is for $m_{t'} =$ 300 GeV. For $b'\to
sZ_L$, $\sin\delta$ hardly drops above the $tW$ threshold, and
rises back to 1 after crossing the $t'W$ threshold.
%
For all cases, we see that near and above the $tW$ threshold, the
absorptive $b' \to tW \to sV^0$ amplitude is optimal for CPV, with
$\sin\delta$ of order 1. The phase difference between
$V_{t's}^*V_{t'b'}$ and $V_{ts}^*V_{tb'}$ provides the CPV weak
phase. We thus foresee that CPV in $b'\to sV^0$ decays is the most
interesting for $b'$ near and above the $tW$ threshold.


What about other $b'$ and $t'$ loop decays? The $b'\to b$, the
$t'\to t$, $c$, and the $t\to c$ transitions all turn out to be
dominated by a single amplitude, and, according to
Eq.~(\ref{eq:cpa}), cannot generate much $A_{\rm CP}$.
For $b'\to b$ decays, $V_{ub}^*V_{ub'}$ and $V_{cb}^*V_{cb'}$ are
rather small, so $V_{tb}^*V_{tb'} \cong -V_{t'b}^*V_{t'b'}$,
hence~\cite{AH06} ${\cal M}_{b'\to b} =
V_{t'b}^*V_{t'b'}\,\Delta(t',t)$.
For $t' \to t$ transitions, $m_d$, $m_s$ and $m_b$ can be taken as
0 as compared to the weak scale, hence ${\cal M}_{t'\to t} =
V_{tb'}V_{t'b'}^*\,\Delta(b',0)$, while one simply changes the CKM
factor to $V_{cb'}V_{t^{(\prime)}b'}^*$ for $t^{(\prime)} \to c$
transitions. Note that the generic loop function $f$ contains
implicit external heavy mass dependence for different transitions.

So, the $b'\to s$ transitions are special. Like $b'\to b$
transitions, they are indeed loop suppressed. With a hint of
possible large CPV effects in $b\to s$
transitions~\cite{CDFsin2betas,HNS,BelleNature}, which is
controlled by the CKM product $V_{t's}^*V_{t'b}$, $b'\to s$ need
not be far suppressed compared to $b'\to b$. From
Eq.~(\ref{eq:Mbptos}), we see that $b'\to s$ transitions are
controlled by $V_{t's}^*V_{t'b'}$, which should not be too
different in length from $V_{cb'}$ that controls the tree level
$b'\to cW$ decay, as $V_{t'b'} \cong 1$ is expected. In the same
way, the $b'\to b$ transitions share similar CKM dependence with
$b'\to \{tW\}^*$, where the $*$ means that below threshold, the
$t$ or the $W$, or both, could be off-shell.

\begin{figure}[t!]
 \vskip-0.1cm
\includegraphics[width=3in]{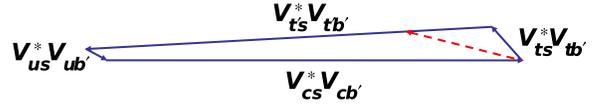} 
 \vskip-0.1cm
 \caption{
 The unitarity quadrangle of Eq.~(\ref{eq:UQbptos}), with tilt of
 $V_{us}^*V_{ub'}$ exaggerated. Dashed line is explained in text.
 \label{fig:Quad}
}
\end{figure}

As a starting point of our numerical study, we shall use the
explicit $4\times 4$ CKM matrix from the second reference of
Ref.~\cite{HNS}, which is an illustration for $m_{t'} = 300$ GeV
from a comprehensive study of $b\to s$, $s\to d$ and $b\to d$ FCNC
and CPV processes. The value for $|V_{cb'}| \sim 0.116$ is rather
sizable. However, if the $\Delta A_{K\pi}$ and $\sin2\Phi_{B_s}$
indications are taken seriously, then choosing a smaller value for
$|V_{t'b}|$ than 0.22 (taken in Ref.~\cite{HNS}), then $|V_{t's}|$
hence $|V_{cb'}|$ would only be larger. The unitarity quadrangle
relevant for $b'\to s$ transitions is
\begin{eqnarray}
 0 &=&  \left(V_{us}^*V_{ub'} + V_{cs}^*V_{cb'}
  + V_{ts}^*V_{tb'} + V_{t's}^*V_{t'b'}\right) 10^{2} e^{-i\,66^\circ}
 \nonumber \\
 &=&  0.63\,e^{-i\,5^\circ} + 11.25
    - 1.20\,e^{-i\,42^\circ} - 11.01\,e^{i\,4^\circ},
 \label{eq:UQbptos}
\end{eqnarray}
and is depicted in Fig.~\ref{fig:Quad}. In discussions, however,
we shall allow variations to illustrate the full range of possible
CPV effects, e.g. the dashed line illustrates the case of a
smaller $V_{cs}^*V_{cb'}$. Note that the quadrangle
in Fig.~\ref{fig:Quad} has the same area as the $b\to s$
quadrangle shown in Refs.~\cite{HNS,Hou08}, but is somewhat
squashed. This disadvantages the $b'\to s$ process for CPV
purposes. To get the largest CPV asymmetry, from
Eq.~(\ref{eq:cpa}) we see that the two interfering amplitudes
better have similar strength.

\begin{figure*}[t!]
\begin{tabular}{ccc}
 \hspace{-0.2cm}
\resizebox{62mm}{!}{\includegraphics{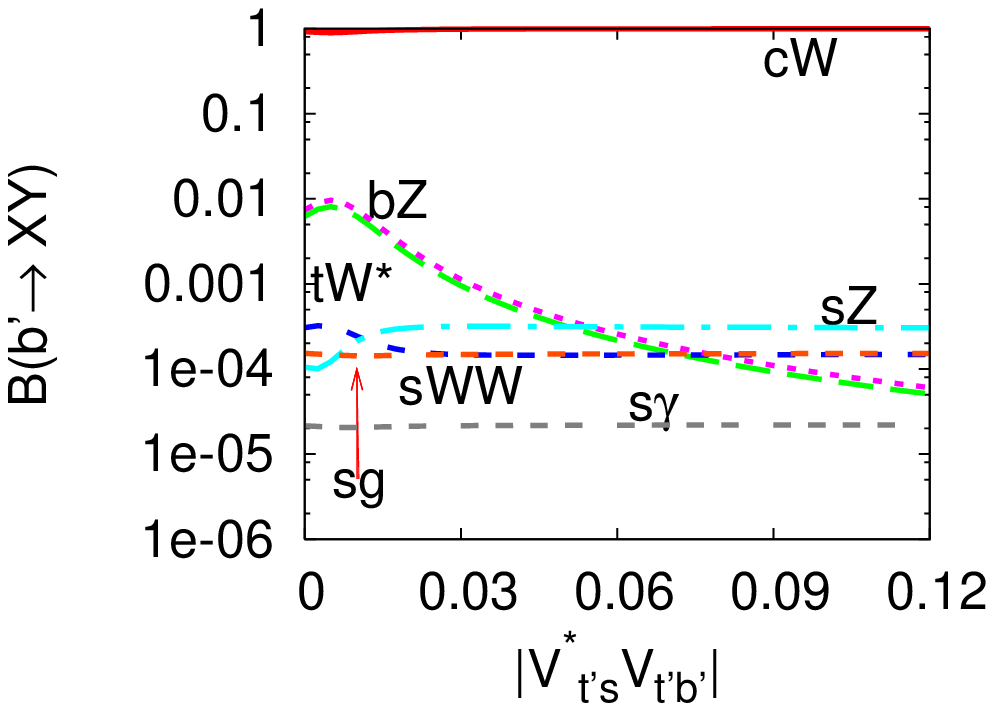}} & \hspace{-1.8cm}
\resizebox{62mm}{!}{\includegraphics{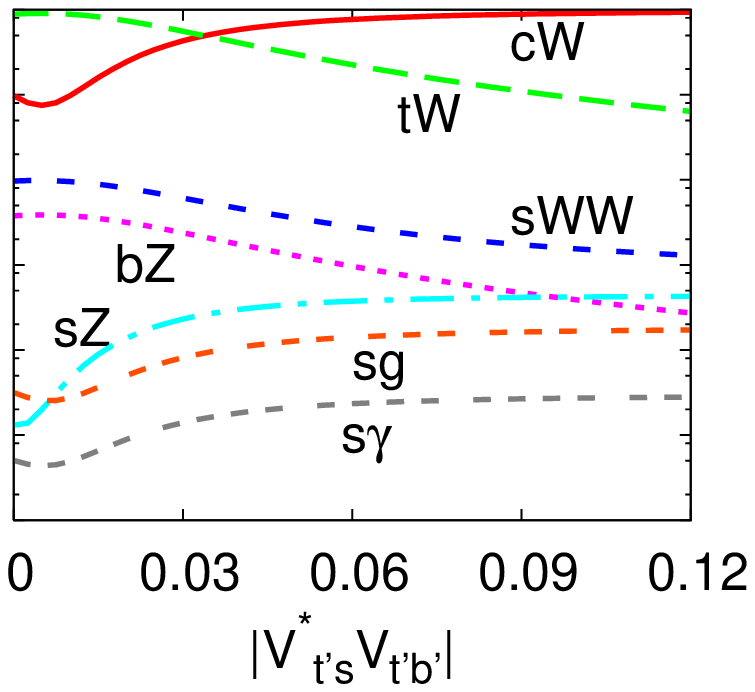}} & \hspace{-2.1cm}
\resizebox{62mm}{!}{\includegraphics{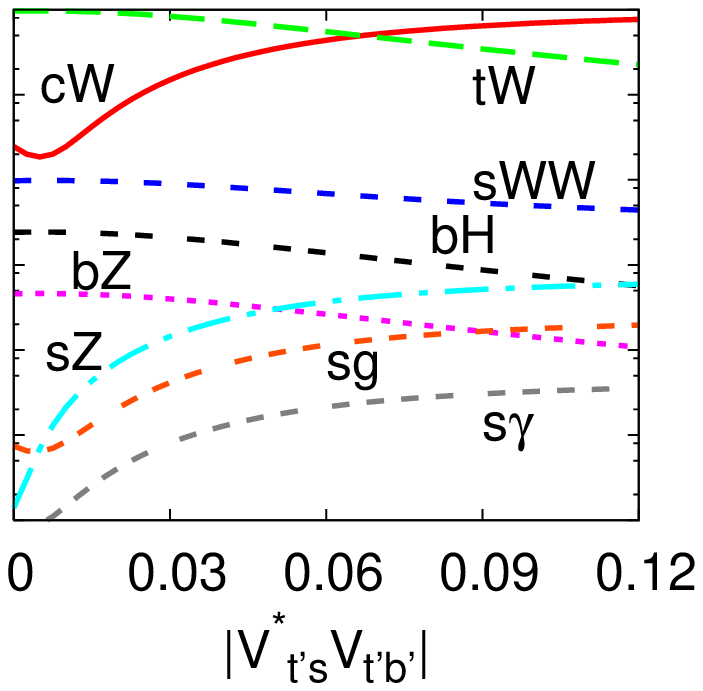}}
\end{tabular}
\caption{Branching ratios of $b'$ decays as a function of
$|V_{t's}^*V_{t'b'}|$ with $\arg (-V_{t's}^*V_{t'b'}) = 70^\circ$,
and $V_{ts}^*V_{tb'} = -0.01\, e^{i\,10^\circ}$, $V_{tb'}=0.1$,
for $m_{b'}=210$ GeV (left), 260 GeV (center) $m_{b'}=340$ GeV
(right). The CKM factors are close to, but not exactly the same as
the numerical example of Ref.~\cite{HNS}, given in
Eq.~(\ref{eq:UQbptos}).}
 \label{fig:BRbp}
\end{figure*}

It is known~\cite{HouStuart} that the intertwining effects of CKM,
kinematic and loop suppressions make $b'$ decays particularly rich
and interesting, especially for $b'$ below $tW$ threshold. The
potential ``cocktail solution"~\cite{AH01} can certainly evade
current CDF search bounds. With the potential indication of large
CPV activity in $b\to s$ transitions, $V_{t's}$ and $V_{t'b}$
could in fact be comparable in strength, further enriching the
$b'$ decay scenario.
We illustrate possible branching fractions of various decay modes
for $b'$ in Fig.~\ref{fig:BRbp}, for the low, near and above $tW$
threshold masses of $m_{b'} =$ 210, 260, 340 GeV.

\begin{figure}[b!]
\begin{tabular}{cc}
\hspace{-.1cm} \resizebox{44mm}{!}{\includegraphics{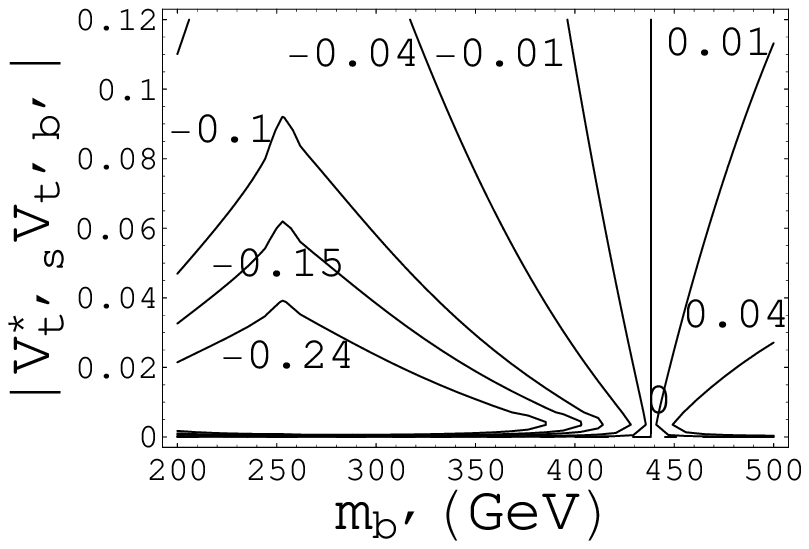}} &
\hspace{-0.9cm} \resizebox{44mm}{!}{\includegraphics{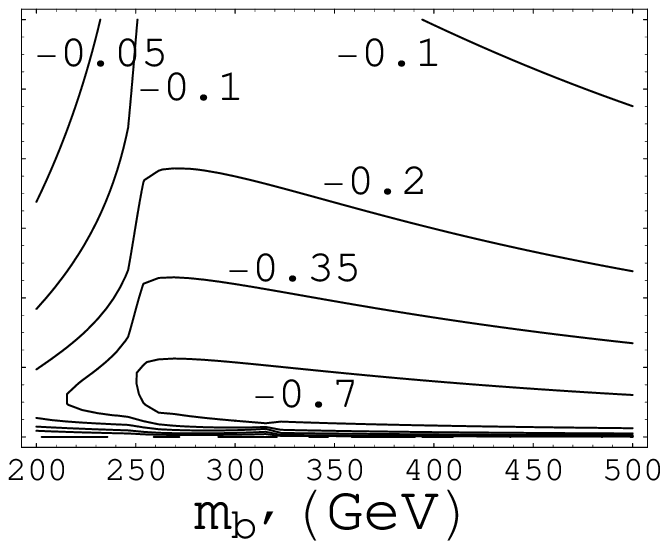}}
\vspace{-0.15cm}
\end{tabular}
\caption{Contour plots of $A_{\rm CP}$ for $b'\to sZ$ (left) and
 $b'\to s\gamma$ (right) in $m_{b'}$--$|V_{t's}^*V_{t'b'}|$
 plane, with $m_{t'}=m_{b'}+50$ GeV, and CKM parameters as in
 Fig.~\ref{fig:BRbp}.}
 \label{fig:ACPmbp}
\end{figure}

The contours of $A_{\rm CP}$ in the plane of $m_{b'}$ and
$|V_{t's}^*V_{t'b'}|$ is plotted in Fig.~\ref{fig:ACPmbp} for
$b'\to sZ$ and $s\gamma$ decays, with $m_{t'}=m_{b'}+50$ GeV. One
sees clearly that the largest $A_{\rm CP}$ occurs around $tW$
threshold, and for $|V_{t's}^*V_{t'b'}|$ approaching
$V_{ts}^*V_{tb'}$ in strength. We remark that $|V_{ts}^*V_{tb'}|
\sim 0.01$ is about as large as it can get. In the limit that all
rotation angles are smaller than $V_{us}$ (Cabibbo angle), which
seems to be the case, $|V_{ts}|$ remains close to the measured
$V_{cb}$ in strength, and cannot be larger than about 0.05, while
$V_{t'b} \sim 0.2$ is the bound from $Z\to b\bar b$ for $m_{t'}
\sim 300$ GeV. For larger $m_{b'}$, $V_{t'b}$ will have to be
smaller. We see from Fig.~\ref{fig:ACPmbp} that maximal $A_{\rm
CP}$ occurs for $|V_{t's}^*V_{t'b'}|$ not far from
$|V_{ts}^*V_{tb'}|$, as can be understood from Eq.~(\ref{eq:cpa}).
However, if Eq.~(\ref{eq:UQbptos}) holds, then $A_{\rm CP}$ is
less than $-10\%$, smaller when away from the $tW$ threshold. Note
that the $CP$ asymmetry flips sign for $m_{b'}$ above 440 GeV or
so, which is due to the change in sign of the $CP$ conserving
phase difference $\sin\delta$, as depicted in
Fig.~\ref{fig:sindelta}.

For the $b'\to s\gamma$ case ($b'\to sg$ is qualitatively
similar), because of gauge invariance, it can only be induced by a
dipole transition, and the $m_t$, $m_{t'}$ dependence is
different. As can be seen from Fig.~\ref{fig:sindelta}, the $CP$
conserving phase difference $\sin\delta$ turns on sharply above
the $tW$ threshold, becoming close to 1 unless the $t'W$ threshold
is encountered (never the case if EWPT is respected). There is no
flip of sign for $\sin\delta$, and we see from
Fig.~\ref{fig:ACPmbp}(b) that rather large $A_{\rm CP}$ can be
attained for $|V_{t's}^*V_{t'b'}| \sim |V_{ts}^*V_{tb'}|$ , even
towards high $b'$ masses. But if Eq.~(\ref{eq:UQbptos}) holds,
then $A_{\rm CP}$ cannot be more than $-15\%$, and the best mass
range is above $tW$ threshold, up to 350 GeV or so.

\begin{figure}[b!]
\begin{tabular}{cc}
\hspace{-.3cm} \resizebox{44mm}{!}{\includegraphics{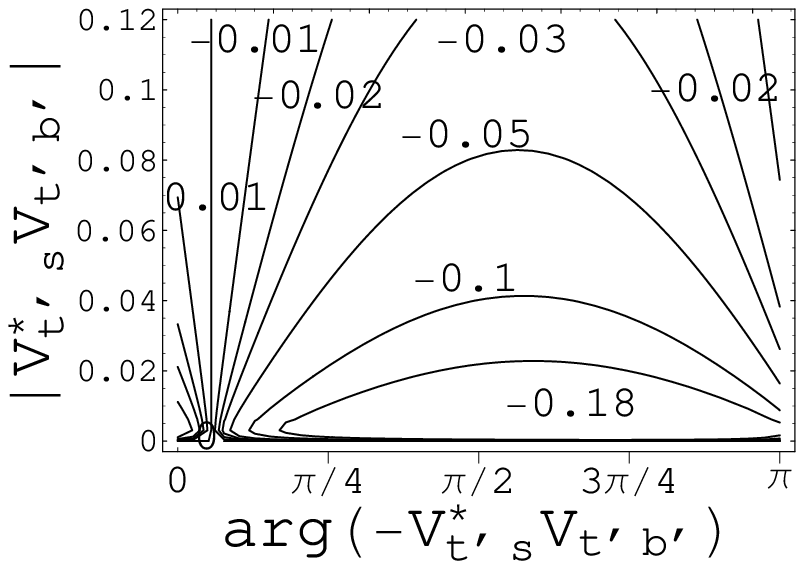}} &
\hspace{-1cm} \resizebox{44mm}{!}{\includegraphics{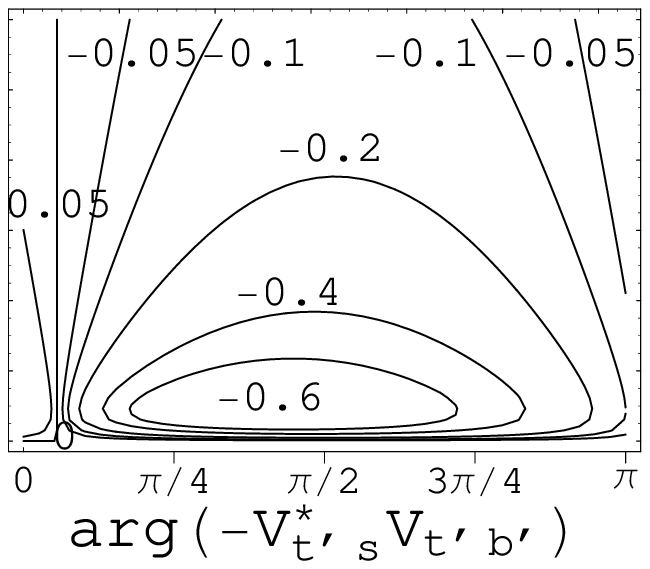}}
\vspace{-0.15cm}
\end{tabular}
\caption{$A_{\rm CP}$ contours for $b'\to sZ$ (left) and
 $b'\to s\gamma$ (right) in $\arg (-V_{t's}^*V_{t'b'})$--$|V_{t's}^*V_{t'b'}|$
 plane, with $m_{b'} =340$ GeV and $m_{t'}=m_{b'}+50$ GeV, and
 $V_{ts}^*V_{tb'} = -0.01\, e^{i\,10^\circ}$, $V_{tb'}=0.1$.}
 \label{fig:ACPphi_sbp}
\end{figure}

So far we have fixed the phase of $V_{t's}^*V_{t'b'}$ to
$70^\circ$, as suggested by Ref.~\cite{HNS}, and given in
Eq.~(\ref{eq:UQbptos}). To illustrate the dependence on this
phase, we plot in Fig.~\ref{fig:ACPphi_sbp} $A_{\rm CP}$ contours
in the $\arg (-V_{t's}^*V_{t'b'})$--$|V_{t's}^*V_{t'b'}|$ plane.
Since $\arg (-V_{t's}^*V_{t'b'}) \sim 70^\circ$ is already very
large, the gain for a CPV phase of $\pi/2$ is not dramatic, but
$A_{\rm CP}$ of course vanishes if this phase turns out to be
small.

\begin{figure}[t!]
\begin{tabular}{cc}
 \vspace{0.2cm}
 \hspace{-.5cm}\resizebox{41.5mm}{!}{\includegraphics{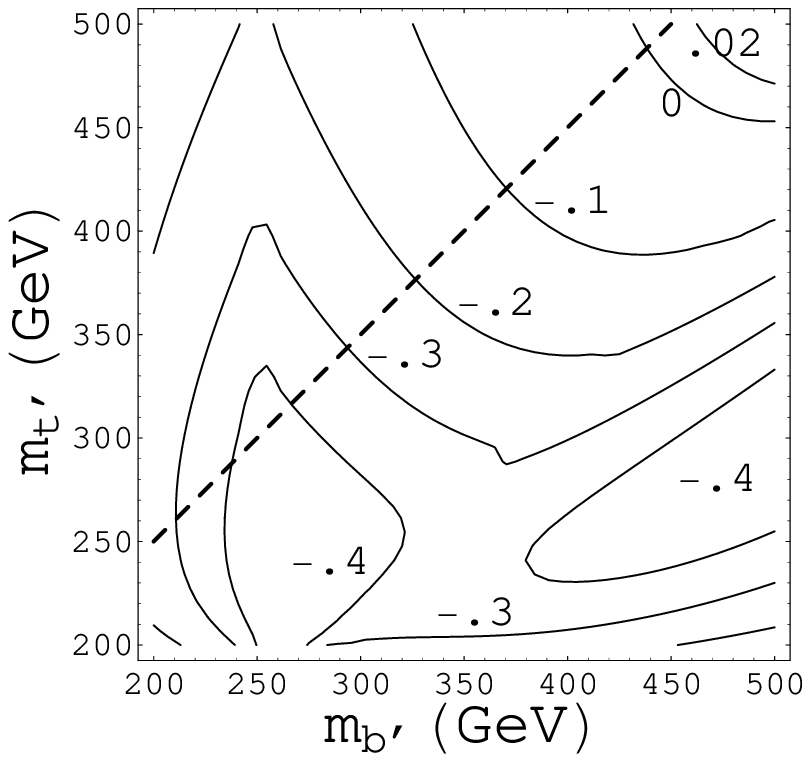}} &
 \hspace{-0.5cm}\resizebox{41.5mm}{!}{\includegraphics{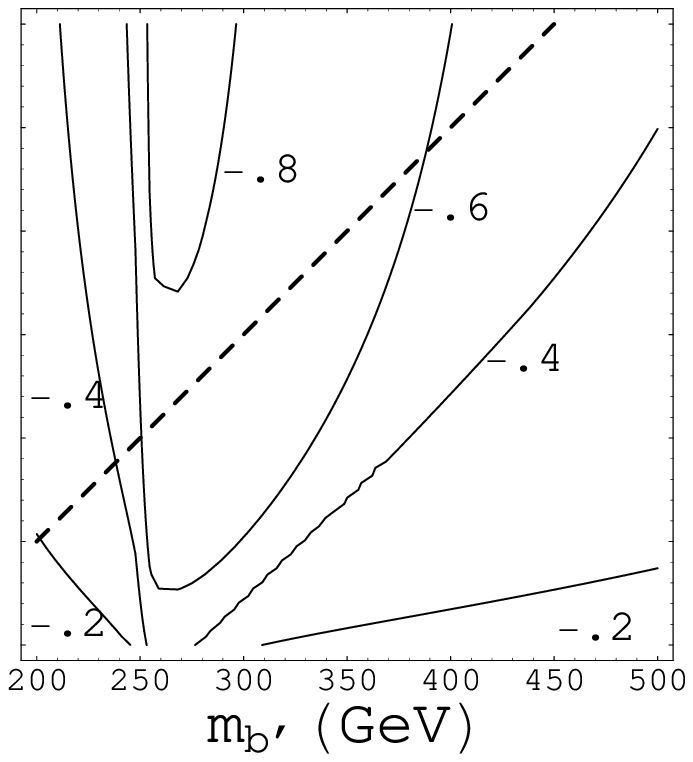}}
 \vspace{-0.2cm}
\end{tabular}
\caption{$A_{\rm CP}$ contours for $b'\to sZ$ (left) and
 $b'\to s\gamma$ (right) in $m_{b'}$--$m_{t'}$ plane, for
 $V_{t's}^*V_{t'b'} \simeq -0.02\, e^{i\,70^\circ}$,
 $V_{ts}^*V_{tb'} \simeq -0.01\, e^{i\,10^\circ}$.
 The dashed line is for $m_{t'} = m_{b'} + 50$ GeV.}
 \label{fig:ACPmbpmtp}
\end{figure}

We have been applying $m_{t'} = m_{b'} + 50$ GeV that respects
electroweak constraints.
%
To get a feeling of the broader behavior, in
Fig.~\ref{fig:ACPmbpmtp} we release the EWPT constraint and plot
$A_{\rm CP}$ contours in the $m_{b'}$--$m_{t'}$ plane. We select
the near optimal (for strength of $A_{\rm CP}$) scenario of
$V_{t's}^*V_{t'b'} \simeq -0.02\, e^{i\,70^\circ}$, with our
nominal $V_{ts}^*V_{tb'} \simeq -0.01\, e^{i\,10^\circ}$. Smaller
values for $|V_{t's}^*V_{t'b'}| \simeq |V_{ts}^*V_{tb'}|$ can give
even larger $CP$ asymmetries. Fig.~\ref{fig:ACPmbpmtp} should be
compared with Fig.~\ref{fig:sindelta}, e.g. for $m_{t'} = m_{b'} +
50$ GeV along the illustrated dashed line (or held fixed at
$m_{t'} =$ 300 or 500 GeV).
We see that, in case of $b'\to sZ$, the largest $CP$ asymmetry is
for $V_{t's}^*V_{t'b'}$ close to $V_{ts}^*V_{tb'}$ in strength,
with large CPV phase difference, and for $m_{t'}$ around $tW$
threshold. For the $b'\to s\gamma$ case, the largest asymmetry
occurs for $m_{b'}$ just above $tW$ threshold, but $m_{t'}$ very
heavy. These masses may not be realistic because of EWPT
constraints, but they illustrate the potential range of CPV for
$b'\to s$ loop transitions.

We note that, to have enhanced $A_{\rm CP}$ for $b'\to s$ decays,
it often may not coincide with large $\sin2\Phi_{B_s}$ in $B_s \to
J/\psi\phi$. For instance, the preference for $V_{t's}^*V_{t'b'}$
close to $V_{ts}^*V_{tb'}$, as illustrated by the red dashed line
in Fig.~\ref{fig:Quad}, $V_{t's}^*$ has practically shrank by 1/4
to 1/5 in length, so $\sin2\Phi_{B_s}$ may not be as strong as the
current Tevatron central value, but would still be very
interesting for LHCb. Likewise, $K_L\to \pi^0\nu\bar\nu$ and CPV
in $D^0$ mixing are open questions. Thus, the impact of a 4th
generation on BSM CPV effects is quite an open question, to be
pursued at multiple fronts.


We see that $A_{\rm CP}(b'\to sZ)$ can in principle go up to
30\%--40\%, especially if $m_{t'}$, $m_{b'} \lesssim 350$ GeV or
so (and $m_{b'} > m_{t'}$ would be better). Larger asymmetries are
possible for $b'\to s\gamma$, and extending to high $m_{b'}$. But
what about the current CDF search bounds? Note that these assume
either $t' \to qW$ (no $b$-tagging) or $b'\to tW$ to be 100\%. As
illustrated in Fig.~\ref{fig:BRbp}, for $m_{b'}$ in the range from
$tW$ threshold to 350 GeV or so, branching ratios depend on
$V_{cb'}$ ($V_{t's}V_{t'b'}$ for $b'\to s$) and $V_{tb'}$
($V_{t'b}V_{t'b'}$ for $b'\to b$) for the $b'\to cW$ and $b'\to
tW$ processes, respectively, and may still evade current bounds.
For example, the smaller value of $V_{t's}^*V_{t'b'}$ that drives
up $A_{\rm CP}$, as illustrated by the red dashed line of
Fig.~\ref{fig:Quad}, implies smaller $b'\to cW$ and $b'\to s$
rates. Of course, $V_{tb'}$ could be smaller as well, but this
could be compensated by a larger $b'$ mass than in
Fig.~\ref{fig:BRbp}(a). What is not sufficiently illustrated in
Fig.~\ref{fig:BRbp} is that $V_{cb'}$ can be suppressed, but with
$V_{t's}^*V_{t'b'}$ less suppressed (we have treated the reverse
case). This is because of CKM unitarity (similar to $V_{td} \neq
0$ in $V_{ub} \to 0$ limit), and again illustrates the potential
richness of multiple decay channels for $b'$, which is best
studied with LHC data in the near future.

Besides $A_{\rm CP}$ and branching fraction, the actual
measurability of CPV in $b'\to s$ decays also requires tagging.
%
Let us take $b'\bar b'$ production cross section $\sim 40$ pb for
$m_{b'} \sim$ 300 GeV with 14 TeV running of LHC, and a typical
$b'\to sZ$ branching ratio of order a few $\times 10^{-4}$. The
tagging efficiency for the other $\bar b'$ (gaining a factor of
two in a $b'\bar b'$ event) is hard to estimate at present, and
would depend on $b'\to cW$ vs $\{tW\}^{(*)}$ fractions. With 100
fb$^{-1}$, we very roughly infer that a 10\% asymmetry may be
reachable with $3\sigma$ statistical significance. The $b'\to
s\gamma$ mode may be more promising. Though typically 1/10 the $b'
\to sZ$ rate, there is no need for $Z\to \ell^+\ell^-$
reconstruction, and the asymmetry could be 50\% or larger. The
$b'\to sg$ would suffer more background for CPV studies because of
lack of distinct signature, and would be less useful. The
situation may be better for an $e^+e^-$ linear collider
environment.

We note that if large $\sin2\Phi_{B_s}$ is observed, hence
something closer to Eq.~(\ref{eq:UQbptos}) or Fig.~\ref{fig:Quad}
is realized, it would imply smaller $A_{\rm CP}$. If
$\sin2\Phi_{B_s}$ is found positive, the $A_{\rm CP}$s discussed
here would flip in sign. Note also that $b'\to sWW$ mode could
also exhibit CPV, but the asymmetry is suppressed by
$\Gamma_t/m_t$ or $\Gamma_W/M_W$ and small.



In conclusion, the best scenario for CPV studies at high energy
collider is for $b'\to s$ decays. 
One would first have to discover the 4th generation, preferably
around 300 GeV or so.
After sorting out the dominant decays, one would have to identify
$b'\to sZ$ or $b'\to s\gamma$ channels, and then tag the other
$\bar b'$.
The study of CPV in $b'$ decays would demand more than 100
fb$^{-1}$ at the LHC. But if measured, one could extract the CPV
phase, since the $CP$ conserving phase is calculable.

\vskip 0.2cm \noindent{\bf Acknowledgement}.\ We thank the
National Science Council of R.O.C. for partial support.


\begin{thebibliography}{99}
\bibitem{Nnu}
  $N_\nu \cong 3$ on the $Z$ peak is
  muted by the observation of neutrino oscillations, which implies a BSM mass scale.
\end{thebibliography}
\end{document}